 \documentclass[iop]{emulateapj-rtx4}

\shorttitle{A new black hole transient near M31*}
\shortauthors{Barnard et al.}

\begin{document}


\title{Discovery of a new black hole transient within 100$''$ (400 pc) of M31*}


\author{R. Barnard, and M. R. Garcia and F. Primini}
\affil{Harvard-Smithsonian Center for Astrophysics, Cambridge MA 02138}
\and
\author{S. S. Murray}
\affil{Johns Hopkins University, Baltimore, Maryland, and CFA}


\begin{abstract}
We identified a new X-ray transient CXOM31 004252.457+411631.17  (T13) in M31 during a 2013 June Chandra observation. This system is particularly exciting because it is located within 100$''$ of M31*; it is thought that this region of the bulge is sufficiently dense to form X-ray binaries dynamicaly, but only systems with black hole accretors and/or short periods are expected to survive. A follow-up  XMM-Newton observation yielded a soft spectrum, well described by a 0.39$\pm$0.02 keV disk blackbody; applying this model to the Chandra observation yielded an observed 0.3--10 keV luminosity peak of 6.2$\pm$0.6$\times 10^{37}$ erg s$^{-1}$ (4.7$\times 10^{36}$ erg s$^{-1}$ in the 2.0--10 keV band). Observing with HST/ACS did not reveal an optical counterpart, but allowed us to place an  upper limit of $B$ $>$26.9, corresponding to an absolute V band magnitude $>$2.0. From the 2--10 keV luminosity and absolute V magnitude, we estimate an orbital period $<$5 h from an empirical relation. Fitting a disk blackbody + blackbody model allows us to reject a neutron star accretor at a 14$\sigma$ level.

\end{abstract}





\section{Introduction}

We have been monitoring the central region of M31 for the last $\sim$13 years with {\em Chandra}, averaging $\sim$1 observation per month, in order to discover X-ray transients. Promising examples are followed up with two {\em HST} ACS observations, the first is taken a few weeks after outburst, and the second observation is taken $\sim$6 months later; this allows us to identify the counterpart via difference imaging \citep[see e.g.][ and  references within]{barnard2012b}. We summarized the results of the first 12 transients (labeled T1--T12) found via this effort in \citet{barnard2012b}. In this work, we study CXOM31 004252.457+411631.17, referred to hearafter as T13.

We discovered T13 in the 2013 June, 5ks Chandra ACIS observation of the M31 center, $\sim$95$''$ from the M31 nucleus (M31*) \citep{barnard13c}. We obtained 92 net source photons, insufficient for spectral modeling; however, the transient was likely to be in one of two spectral states, and we estimated the luminosity for both states. Assuming a  1 keV disk blackbody model with line-of sight absorption equivalent to 7$\times 10^{20}$ H atom cm$^2$ yielded a 0.3--10 keV luminosity of 6.1$\pm$0.6$\times 10^{37}$ erg s$^{-1}$ \citep{barnard13c}; this corresponds to a black hole binary in a thermally dominated state \citep{remillard06}. A power law emission model with spectral index 1.7 yielded a 0.3--10 keV luminosity of 8.9$\pm$0.9$\times 10^{37}$ erg s$^{-1}$ \citep{barnard13c}; this corresponded to a black hole binary in its low state \citep{remillard06}. Hence, the two possible spectral states for T13 yielded rather similar 0.3--10 keV luminosities. We found that the locale of  T13 was serendipitously observed in one of our earlier HST observations, meaning that we only required one new HST observation.

X-ray transients within the central region of M31 are most likely to contain a low mass secondary, as the majority of stars there are old.   Low mass X-ray binaries may be transient X-ray sources due to instabilities in their accretion disks; the disk has two stable phases (hot and cold), and an unstable intermediate phase--- matter accumulates in the disk in the cold phase, and is rapidly dumped onto the compact object in the hot phase \citep[see e.g.][]{lasota2001}. However, the X-rays produced by accretion from the hot disk prevent the disk from cooling; the X-ray luminosity decays exponentially if the whole disk is ionized, and linearly if only part of the disk is ionized \citep{king98}.

\citet{vp94} found an empirical relation between the  ratio of  X-ray and optical luminosities of Galactic X-ray binaries  and their orbital periods, suggestive that the optical emission is dominated by reprocessed X-rays in the disk; this relation holds over a 10 magnitude range in optical luminosity, and appears to be insensitive to inclination. Their chosen X-ray band was 2--10 keV. For an irradiated accretion disk with radius $a$, X-ray luminosity $L_{\rm X}$, optical luminosity $L_{\rm opt}$, and temperature $T$, T$^4$ $\propto$ $L_{\rm X}$/$a^2$, while the surface brightness of the disk, $S$, $\propto$ T$^2$ for typical X-ray binaries \citep{vp94}. Since $L_{\rm opt}$ $\propto$ $S.a^2$,  $L_{\rm opt}$ $\propto$ $L_{\rm X}^{1/2} a$; also $a$ $\propto$ $P^{2/3}_{\rm orb}$, where $P_{\rm orb}$ is the orbital period. 

\citet{vp94} defined $\Sigma$ = $\left(L_{\rm X}/L_{\rm EDD}\right)^{1/2}\left(P_{\rm orb}/1 {\rm hr}\right)^{2/3}$, choosing   $L_{\rm EDD}$ = 2.5$\times 10^{38}$ erg s$^{-1}$ as a normalizing constant, and found 
\begin{equation}
M_{\rm V} = 1.57(\pm0.24) - 2.27(\pm 0.32) \log \Sigma.
\end{equation}
However, \citet{vp94} sampled a mixture of neutron star and black hole binaries, in various spectral states. A cleaner sample was obtained by A. Moss et al. (2013, in prep), who used only black hole transients at the peaks of their outburst, and found
\begin{equation}
M_{\rm V} = 0.84(\pm0.30) - 2.36(\pm0.30) \log \Sigma.
\end{equation}
We note that these two relations only differ significantly in normalization, caused by black hole X-ray binaries having larger disks than neutron star binaries with the same period. We have period estimates for 12 M31 transients (T1--T12) observed by {\em Chandra} and {\em HST} \citep{barnard2012b}.

T13 is particularly interesting because of its close proximity to M31*. \citet{voss07} found an excess of XBs within $\sim$100$''$ of M31* when they compared the surface density  of XBs with the stellar mass as  traced by the K band light. They proposed that these excess binaries could be formed dynamically in the bulge of M31, as seen in globular clusters. They noted that the stellar velocities in the bulge are considerably higher than in globular clusters, and concluded that surviving dynamically-formed XBs in the bulge would likely have a black hole primary and / or a very short period. T13 could be just such a system, meaning that its existence could provide strong support for the theory proposed by \citet{voss07}.

We have observed T13 twice with Chandra, and once more with HST. Furthermore, we were granted an $\sim$8 ks XMM-Newton observation as part of the target of opportunity (TOO) program. In this paper we present our analysis of the Chandra, HST, and XMM-Newton data.

\section{Observations and data reduction}

We  observed T13 in two 5 ks Chandra ACIS observations, on 2013 June 2 (C1, ObsID 15324, PI Murray), and 2013 July 24 (C2, ObsID 15326, PI Murray). We observed T13 in outburst for 4.8 ks with HST/ACS WFC using the F435W filter on 2013 June 18 (H2, jc6b01010, PI Barnard); we also serendipitously observed T13 in quiescence during our 4.7 ks, 2007 January 10 HST/ACIS observation in the same configuration (H1, j9ju06010, PI Garcia). Additionally, we were granted an $\sim$8 ks XMM-Newton observation on 2013 July (X1, 0727960401, PI N. Schartel) using the Thin filter. Our analysis of the Chandra and HST data follows the procedures outlined in \citet{barnard13b}.

\subsection{X-ray Analysis}
We extracted spectra from our X-ray observations of T13 using  the appropriate mission-specific software suites: CIAO v.4.4 for C1--C2, and {\em XMM-Newton} SAS v.13.0.0 for X1. We used the CALDB version distributed with CIAO. Spectral analysis was performed with XSPEC v12.8.0m.

\subsubsection{{\em Chandra} analysis}
For each of our {\em Chandra} ACIS observations, we extracted source and background 0.3--7.0 keV spectra from circular regions with 2.5$''$ radii. We then created the appropriate response files (response matrix and ancilary response file). The source spectra were grouped to give a minimum of 20 counts per bin.

\subsubsection{{\em XMM-Newton} analysis}
 {\em XMM-Newton} observations often experience intervals of greatly increased background levels. We searched for such flaring intervals by creating a lightcurve  using the expression ``(PATTERN==0)\&\&(PI in [10000:12000])\&\&(FLAG==0)'' and 100 s bins; this observation was flare-free.

We extracted a 0.3--10 keV source spectrum from a circular region with 240 pixel radius, to avoid contamination from a nearby source. A circular background region was chosen to be near the source, on the same chip, and at a similar off-axis angle; its radius was 550 pixels. These spectra were filtered by the expression ``(PATTERN$<$=4)\&\&(FLAG==0)''. We also obtained a corresponding response files. The source spectrum was grouped to a minimum of 20 counts per bin.

\subsection{Locating the X-ray source}
We used 27 X-ray bright globular clusters (GCs) to register a combined $\sim$300 ks  ACIS image (supplied by Z. Li) to the B band Field 5 image of M31 provided by the Local Galaxy group Survey (LGS) \citep{massey06}. We used {\sc pc-iraf} v2.14.1 to perform the registration, following the same procedure as described in \citet{barnard2012b}. 
This  yielded  1$\sigma$ position uncertainties of  0.11$''$ in R.A., and 0.09$''$ in Dec \citep{barnard2012b}.

 Similarly, we registered C1 to the merged {\em Chandra} image using a selection of bright X-ray sources. The final uncertainties in the X-ray position of T3 combine the position uncertainties in the X-ray image, the uncertainties in registering Observation C1 to the merged {\em Chandra} image, and the uncertainties in registering the merged {\em Chandra} image to the M31 Field 5 LGS image. The uncertainties in registering the {\em HST} observation were negligible.

\subsection{Optical analysis}

All optical analysis was performed with {\sc pc-iraf} Revision 2.14.1, except where noted. Each {\em HST} observation included four flat-fielded (FLT) images, and one drizzled  (DRZ) image. 
  We used the DRZ images from H1 and H2 to create a difference image; however, we  used the H2 FLT images for our aperture photometry because the software used  ({\sc daophot}) prefers images in that format. 

\subsubsection{Creating a difference image}

 We reprojected the  H2 DRZ  image  into the coordinates of the H1  DRZ image, to produce an accurate difference image;   see \citet{barnard13b} for further details. To do this, we first registered the H1 and H2  images to the LGS Field 5 image, using unsaturated stars that were close to the target; the uncertainties were negligible. Then, we reoriented  the H2 image to match  the H1 image.   The difference image was produced by subtracting H1 from H2. 

\subsubsection{Measuring the optical counterpart}

For H2  we used the {\sc daophot} package released with {\sc iraf} to obtain the net source counts in the FLT images, for a total of $C_{\rm tot}$ counts over $T$ seconds. 
 We converted this  to Vega $B$ magnitude via
\begin{equation}\label{4s}
B \simeq -2.5 \log\left[ C_{\rm tot}/T \right] + 25.77,
\end{equation} 
 following \citet{barnard13b}. We can convert from $B$ magnitude to $M_{\rm V}$ via
\begin{equation}\label{conv}
 M_{\rm V} = B + 0.09   -   N_{\rm H}\times\left(1+1/3\right)/1.8\times10^{21}  -    24.47, 
\end{equation}
where $N_{\rm H}$ is the line of sight absorption; 
\citep[see][and references within]{barnard2012b}.

\begin{figure}[!ht]
\epsscale{1}
\plotone{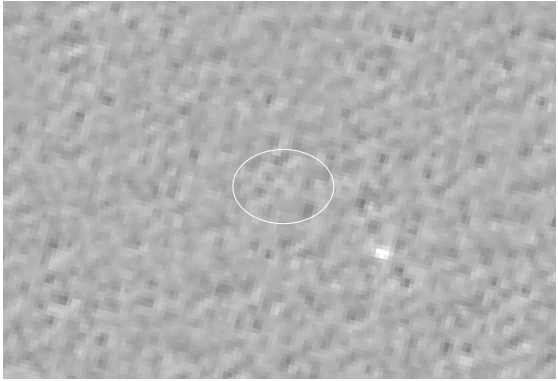}
\caption{A portion of the difference image obtained by subtracting the 2007 HST ACS/WFC F450W image (H1) from the 2013 HST ACS/WFC F450W image (H2). White objects were brighter in 2013 than in 2007. The ellipse represents the 3$\sigma$ uncertainties in the X-ray position of T3: 0.57$''$ in R.A. and 0.42$''$ in Decl. No counterpart was observed.  }\label{diffim}
\end{figure}

\section{Results}

\subsection{Searching for an optical counterpart}

The X-ray position of T13 included final  1$\sigma$ uncertainties of 0.19$''$ in R.A. and 0.14$''$ in Dec. Figure~\ref{diffim} shows a detail of the difference image overlaid with an ellipse that represents the 3$\sigma$ uncertainties in the X-ray position for T13; white objects are brighter during H2 (during outburst) than in H1 (when T13 was in quiescence). We found no evidence for an optical counterpart.

Instead, we calculated the 4$\sigma$ upper limit to the B magnitude of the counterpart by summing the total counts within a circle with 3 pixel radius for each FLT image, and dividing by the total exposure time; this radius corresponds to $\sim$3 times the FWHM for the ACS. We obtained $\sim$180,000 photons over $\sim$4.8 ks. The 4$\sigma$ upper limit was estimated to be 4$C_{\rm Tot}^{0.5}/T$, and   Equation~\ref{4s} yielded $B$ $>$ 26.9 at the 4$\sigma$ level.

\subsection{The XMM-Newton spectrum}
\begin{figure}[!ht]
\epsscale{1}
\plotone{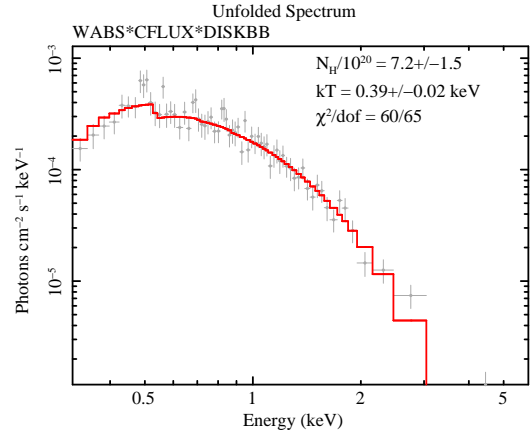}
\caption{The $\sim$0.3--5 keV XMM-Newton pn spectrum of T13, fitted with an absorbed disk blackbody model. }\label{dbspec}
\end{figure}

We obtained 1455 net source counts from the pn spectrum in X1. This spectrum was thermally dominated, and well described by a 0.39$\pm$0.02  keV disk blackbody, with line-of-sight absorption equivalent to 7.2$\pm$1.5$\times 10^{20}$ H atom cm$^{-2}$, and $\chi^2$/dof = 60/65. There was very little power above $\sim$5 keV;  the $\sim$0.3--5 keV spectrum is shown in Figure~\ref{dbspec}.

\begin{figure}[!ht]
\epsscale{1}
\plotone{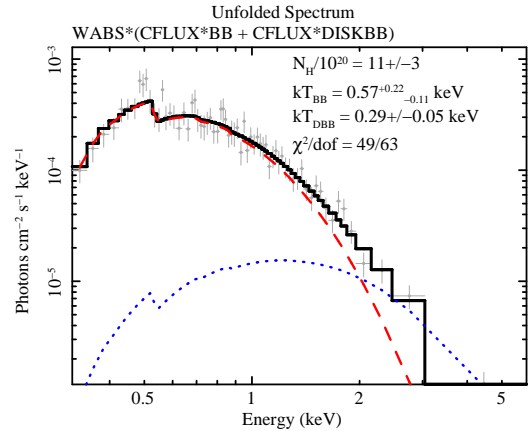}
\caption{The $\sim$0.3--5 keV XMM-Newton pn spectrum of T3 fitted with the double thermal (disk blackbody + blackbody) emission model. The disk blackbody and blackbody components are represented by dashed and dotted lines, respectively; the solid line represents the combined emission.}\label{2cspec}
\end{figure}

We also applied a double thermal emission model (disk blackbody + blackbody), for comparison with the soft states observed in neutron star binaries studied by Lin et al. (2007, 2009, 2012). The best fit yielded a 0.29$\pm$0.05 keV disk blackbody plus a 0.57$^{+0.22}_{-0.11}$ keV blackbody, absorbed by 1.1$\pm$0.3$\times 10^{21}$ H atom cm$^{-2}$; this fit is show in Figure~\ref{2cspec}.  We have used the disk blackbody component in this double thermal model  to identify 35 black hole candidates (BHCs) in M31; neutron star  XBs exhibit soft spectra at luminosities $\ga$3$\times 10^{37}$ erg s$^{-1}$, with disk blackbody temperatures $>$1 keV in all cases, while all 35 BHCs inhabited a separate parameter space \citep{barnard13}. We note that our measured disk blackbody temperature for T13 is 14$\sigma$ below the limit for luminous neutron star XBs, and reject a neutron star  accretor. 

 The 0.3--10 keV unabsorbed luminosity for T13 was 4.0$\pm$0.3$\times10^{37}$ erg s$^{-1}$ in  X1. Therefore the X-ray to optical flux ratio  was $>$1250, ruling out a foreground star. We conclude that T13 contains a black hole. 

 \subsection{The X-ray lightcurve of T13}
If we assume the best fit disk blackbody model for X1, then observations C1 and C2 exhibited 0.3--10 keV luminosities of 6.2$\pm$0.6$\times 10^{37}$ and 1.30$\pm$0.16$\times 10^{37}$ erg s$^{-1}$ respectively. We note that the luminosity for C1 is consistent with the luminosity that we previously obtained by assuming a 1 keV disk blackbody \citep{barnard13c}; hence the luminosity appears fairly insensitive to the disk blackbody temperature.  Unfortunately, M31 was unobservable by Chandra in the $\sim$3 months before C1, hence the start time of the outburst is uncertain; the peak is likely to have been unobserved. 

Transient decays are expected to be exponential while the whole disk is illuminated, or linear when the disk is only partially illuminated \citep{king98}. \citet{shahbaz98} examined Galactic NS and BH transients with known orbital periods and peak 0.4--10 keV  fluxes and established that they followed this  expectation. From Figure 1 of \citet{shahbaz98}, we only expect linear decay for systems with orbital periods $\ga$30 hr, assuming a 10 $M_\odot$ primary, and a peak luminosity of $\sim$6$\times 10^{37}$ erg s$^{-1}$. We find a period $<$5 hr for T13 (see next section); hence, we expect the decay to be exponential for a $\sim10 M_\odot$ BH.

\begin{figure}[!ht]
\epsscale{1}
\plotone{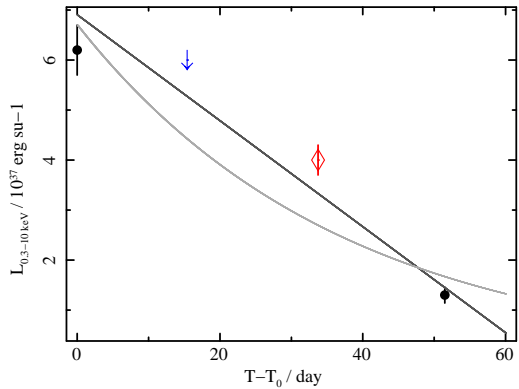}
\caption{The 0.3--10 keV luminosity lightcurve for T13; C1 and C2 are represented by circles, and X1 is represented by a diamond. A downward arrow represents the observation  of H2. We present the best fit linear and exponential fits. Time is quoted relative to C1.}\label{toolc}
\end{figure}

We present the 0.3--10 keV unabsorbed luminosity lightcurve for T13 in Figure~\ref{toolc}; the Chandra observations are represented by circles, while the XMM-Newton observation is represented by a diamond. A downward arrow indicates the time that H2 occurred.  Time is quoted relative to C1. We present the best linear and exponential fits; the lightcurve actually looks more linear than exponential, which may point to an accretor that is significantly more massive than 10 $M_\odot$. If the curve is basically exponential, then it has an e-folding time of $\sim$37 days. 

\subsection{Estimating the orbital period}
Since T13 appears to be a black hole transient in its high state, we estimated its orbital period using Equation 2; this relation requires the 2.0--10 keV luminosity. C1 was the closest observation to H2, with a separation of 15.4 days. Since the optical counterparts of X-ray transients tend to decay 2.2 times more slowly \citep{chen97}, we expect our measured $M_{\rm V}$ to be 17\% fainter than at the time of C1. 

Assuming a 0.39 keV disk blackbody, and $N_{\rm H}$ = 7.2$\times 10^{20}$ atom cm$^{-2}$, the 2.0--10 keV luminosity of T13 during C1 was 4.7$\times 10^{36}$ erg s$^{-1}$. We estimate an orbital period $<$5 h for T13 from our 4$\sigma$ upper limit to the B band luminosity, not including uncertainties in Equation 2; we used the 4$\sigma$ limit to be consistent with the optical processing software. If we take the 3$\sigma$ B band limit, $B$ $>$ 27.2, $M_{\rm V}$ $>$ 2.3, and the orbital period is $<$3 h. T13 appears to be the sort of system predicted by \citet{voss07}.

%


\section{Discussion and conclusions}
We have analyzed Chandra, HST and XMM-Newton observations of a new X-ray transient (T13) that appeared within 100$''$ of M31* in 2013 June. We obtained a 4$\sigma$ upper limit to the B band luminosity of $B$ $>$26.9; this translates to an absolute V band magnitude $>$2.0.

The Chandra spectra did not yield sufficient counts for spectral fitting; however, the XMM-Newton pn spectrum yielded 1455 net source counts. The spectrum was well described by a 0.39$\pm$0.02 keV disk blackbody. We also fitted a disk blackbody + blackbody model, for comparison with the complete range of soft neutron star  XB spectra observed by Lin et al. (2007, 2009, 2012); we rejected a neutron star  accretor at a 14$\sigma$ level.

We estimate an orbital period $<$5 h, from an empirical relation that relates the X-ray to optical flux ratio and the orbital period. T13 appears to be just the type of system predicted by \citet{voss07}, lending support to the theory that the bulge is sufficiently dense to form some X-ray binaries dynamically.

\section*{Acknowledgments}
We thank the anonymous referee for their constructive comments. We thank Z. Li for merging the {\em Chandra} data. This research
has made use of data obtained from the {\em Chandra} satellite,
and software provided by the {\em Chandra} X-Ray Center (CXC).
We also include analysis of data from {\em XMM-Newton}, an ESA
science mission with instruments and contributions directly
funded by ESA member states and the US (NASA); we are very grateful to Norbert Schartel and the XMM-Newton team for granting our TOO observation. R.B. is
funded by {\em Chandra} grants GO2-13106X and GO1-12109X,
along with {\em HST} grants GO-11833 and GO-12014. M.R.G. and S.S.M are partially supported by NASA contract  NAS8-03060.





{\it Facilities:} \facility{\em CXO (ACIS)} \facility{{\em HST} (ACS)} \facility{XMM (pn)} 








\end{document}